\documentclass[letterpaper,twocolumn,english,aps,prl,preprintnumbers,groupaddress,twocolumn]{revtex4}
\usepackage[T1]{fontenc}
\usepackage[latin9]{inputenc}
\usepackage{float}
\usepackage{units}
\usepackage{bm}
\usepackage{amsmath}
\usepackage{graphicx}
\usepackage{amssymb}

\makeatletter

\floatstyle{ruled}
\newfloat{algorithm}{tbp}{loa}
\floatname{algorithm}{Algorithm}

\@ifundefined{textcolor}{}
{%
 \definecolor{BLACK}{gray}{0}
 \definecolor{WHITE}{gray}{1}
 \definecolor{RED}{rgb}{1,0,0}
 \definecolor{GREEN}{rgb}{0,1,0}
 \definecolor{BLUE}{rgb}{0,0,1}
 \definecolor{CYAN}{cmyk}{1,0,0,0}
 \definecolor{MAGENTA}{cmyk}{0,1,0,0}
 \definecolor{YELLOW}{cmyk}{0,0,1,0}
 }

\@ifundefined{definecolor}
 {\@ifundefined{definecolor}
 {\@ifundefined{definecolor}
 {\usepackage{color}}{}
}{}
}{}

\makeatother
\usepackage{babel}
\makeatother
\usepackage{babel}
\usepackage{algorithm}
\usepackage{algpseudocode}

\makeatother

\usepackage{babel}

\begin{document}

\title{Linear Scaling Solution of the Time-Dependent Self-Consistent-Field
Equations}

\author{Matt Challacombe}

\email{mchalla@lanl.gov}

\affiliation{Theoretical Division, Los Alamos National Laboratory, Los Alamos,
New Mexico 87545}
\begin{abstract}
\textbf{A new approach to solving the Time-Dependent Self-Consistent-Field
equations is developed based on the double quotient formulation of
Tsiper {[}J. Phys. B, 34 L401 (2001){]}. Dual channel, quasi-independent
non-linear optimization of these quotients is found to yield convergence
rates approaching those of the best case (single channel) Tamm-Dancoff
approximation. This formulation is variational with respect to matrix
truncation, admitting linear scaling solution of the matrix-eigenvalue
problem, which is demonstrated for bulk excitons in the polyphenylene
vinylene oligomer and the (4,3) carbon nanotube segment.} 
\end{abstract}

\preprint{\texttt{LA-UR }09-06104 }


\maketitle
The Time-Dependent Self-Consistent-Field equations together with models
that include some portion of the Hartree-Fock (HF) exchange admit
control over the range of self-interaction in the optical response
\cite{citeulike:4792154,JSong09,KIgumenshchev07,RMagyar07}, and are
related to new models of electron correlation based on the Random
Phase Approximation (RPA) \cite{DLu09,FFurche08,JHarl09,JToulouse09}.
Solving the TD-SCF equations is challenging due to an unconventional
\textit{J}-symmetric structure of the naive molecular orbital (MO)
representation, 

\begin{equation}
\begin{pmatrix}\bm{\mathbb{A}} & \bm{\mathbf{\mathbb{B}}}\\
-\mathbb{\bm{B}} & \bm{-\mathbb{A}}\end{pmatrix}\binom{\vec{X}}{\vec{Y}}=\omega\:\binom{\vec{X}}{\vec{Y}}\,,\label{eq:RPA}\end{equation}
where \textbf{\textit{$\mathbb{A}$}} and \textbf{\textit{$\mathbb{B}$}}
are Hermitian blocks corresponding to $4^{th}$ order tensors spanning
transitions between occupied and virtual sub-spaces, $\omega$ is
the real excitation energy and $\vec{v}=\binom{\vec{X}}{\vec{Y}}$
is the corresponding transition density. By construction, the MO representation
allows strict separation between the dyadic particle-hole (\emph{ph})
and hole-particle (\emph{hp}) solutions, $\vec{X}$ and $\vec{Y}$,
for which specialized algorithms exist. Nevertheless, convergence
of the naive $J$-symmetric problem is typically much slower than
the corresponding Hermitian Tamm-Dancoff approximation (TDA), $\mathbb{A}\vec{X}=\omega\vec{X}$,
which is of reduced dimensionality in the MO representation. 

Several TD-SCF eigensolvers are based on the oscillator picture $\begin{pmatrix}0 & \mathbb{\bm{K}}\\
\boldsymbol{\mathbb{T}} & 0\end{pmatrix}\binom{\vec{p}}{\vec{q}}=\omega\:\binom{\vec{p}}{\vec{q}},$ with $\boldsymbol{\mathbb{K}=\mathbb{A}+\mathbb{B}}$ and $\boldsymbol{\mathbb{T}=\mathbb{A}-\mathbb{B}}$
the Hermitian potential and kinetic matrices, and the dual $\left\{ \vec{p},\vec{q\,}\right\} =\left\{ \vec{X}-\vec{Y},\vec{\, X}+\vec{Y}\right\} $
corresponding to position and momentum. This picture avoids the imbalance
$\left\Vert \vec{X}\right\Vert \gg\left\Vert \vec{Y}\right\Vert $
whilst admitting conventional solutions based on the Hermitian matrix
$\mathbb{G}=\mathbb{K}\cdot\mathbb{T},$ as shown by Tamara and Udagawa
\cite{TTamura64} and extended by Narita and Shibuya with second order
optimization of the quotient $\omega^{2}\left[\vec{p},\vec{q}\right]=\vec{q}\cdot\mathbb{G}\cdot\vec{p}/\left|\vec{p}\cdot\vec{q}\right|$
\cite{SNarita92}. More recently, Tsiper considered the quotients

\begin{equation}
{\normalcolor \omega\left[\vec{p},\vec{q}\right]}=\frac{\vec{p}\,\cdotp\boldsymbol{\mathbb{K}}\,\cdotp\vec{p}}{2\,\left|\vec{p}\cdotp\vec{q}\right|}+\frac{q\,\cdotp\boldsymbol{\mathbb{T}}\,\cdotp\vec{q}\:}{2\,\left|\vec{p}\cdotp\vec{q}\right|}\;,\label{eq:Tsiper}\end{equation}
and developed a corresponding dual channel Lanczos solver. Subspace
solvers in this dual representation have recently been surveyed by
Tretiak, Isborne, Niklasson and Challacombe (TINC) \cite{STretiak09},
with comparative results for semi-empirical models.

Another challenge is dimensionality and scaling. Writing Eq.~(\ref{eq:RPA})
in the general form $\mathbf{\mathbb{L}}\cdot\vec{v}=\omega\vec{\, v}$,
admitting arbitrary representation, the superoperator matrix $\mathbb{L}$
is a $\sim N^{2}\times N^{2}$ tetradic, with \emph{N} the number
of basis functions, assumed proportional to system size. In practice
the action of $\mathbb{\mathbf{\mathbb{L}}}$ onto $\vec{v}$ is carried
out implicitly as $\boldsymbol{L}[\boldsymbol{v}]=[\bm{F},\bm{\, v}]+[\bm{G}[\boldsymbol{v}],\,\bm{P}]\,,$
using an existing framework for construction of the effective Hamiltonian
(Fockian) $\boldsymbol{F}$, where $\bm{P}$ is the one-particle reduced
density matrix, $\bm{G}$ is a screening operator involving Coulomb,
exchange and/or exchange-correlation terms and the correspondence
between superoperator and functional notation is given by a tensorial
mapping between diadic and matrix, $\vec{v}_{{\scriptscriptstyle N^{2}\times1}}{\bf \Leftrightarrow v}_{{\scriptscriptstyle N\times N}}$.

Recent efforts have focused on addressing the problem of dimensionality
by employing linear scaling methods that reduce the cost of $\boldsymbol{L}[\cdot]$
within Density Functional Theory (DFT) to ${\cal O}(N)$. However,
this remains an open problem for the Hartree-Fock (HF) exchange, an
ingredient in models that account for charge transfer in the dynamic
and static response, including the Random Phase Approximation (RPA)
at the pure HF level of theory. Likewise, scaling of the TD-SCF eigenproblem
remains formidable due to associated costs of linear algebra, even
when using powerful Krylov subspace methods. Underscoring this challenge,
one of the most successful approaches to linear scaling TD-DFT avoids
the matrix eigenproblem entirely through explicit time-evolution \cite{Yokojima99,Yam2003}.

Linear scaling matrix methods exploit quantum locality, manifest in
approximate exponential decay of matrix elements expressed in a well
posed, local basis; with the dropping of small elements below a threshold,
$\tau_{\mathrm{mtx}}$, this decay leads to sparse matrices and $\mathcal{O}(N)$
complexity at the forfeit of full precision \cite{MChallacombe99,MChallacombe00B,ANiklasson03}.
Likewise, linear scaling methods for computing the HF exchange employ
an advanced form of direct SCF, exploiting this decay in the rigorous
screening of small exchange interactions bellow the two-electron integral
threshold $\tau_{\mathrm{2e}}$ \cite{ESchwegler97}. The consequence
of these linear scaling approximations is an inexact linear algebra
that challenges Krylov solvers due to nested error accumulation, a
subject of recent formal interest \cite{Simoncini05,Simoncini07}.
Consistent with this view, TINC found that matrix perturbation (a
truncation proxy) disrupts convergence of Krylov solvers with slow
convergence, \emph{i.e.} Lanczos and Arnoldi for the RPA, but has
less impact on solvers with rapid convergence, \emph{i.e.} generically
for the TDA or Davidson for the RPA. Relative to semi-empirical Hamiltonians,
the impact of incompleteness on subspace iteration may be amplified
with first principles models and large basis sets (ill-conditioning). 

\begin{figure}[h]
\includegraphics[width=3.3in]{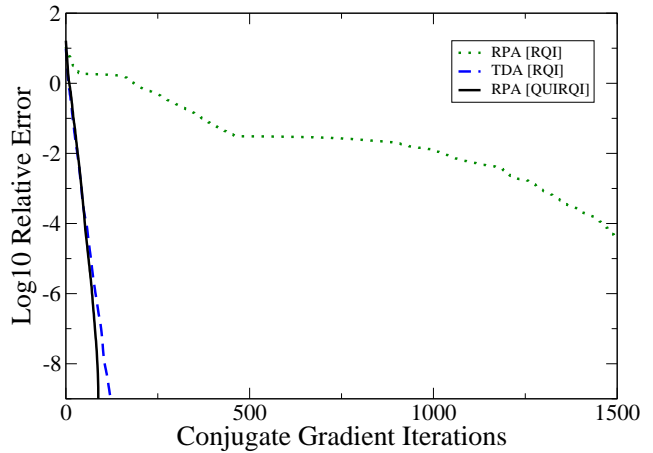}

\caption{Convergence of RHF/3-21G TDA and RPA with the RQI and QUIRQI algorithms
for linear decaene (C$_{10}$H$_{2}$). Calculations were started
from the same random guess, and tight numerical thresholds were used
throughout. In the representation independent scheme, the cost per
iteration is the same for TDA and RPA. \label{fig:RPA_VS_TDA}}

\end{figure}

An alternative is Rayleigh Quotient Iteration (RQI), which poses the
eigenproblem as non-linear optimization and is variational with respect
to matrix perturbation. Narita and Shibuya \cite{SNarita92} considered
optimization of the quotient $\omega^{2}\left[\vec{p},\vec{q}\right]$
with second order methods, but these are beyond the capabilities of
current linear scaling technologies and also, convergence is disadvantaged
by a power of $\nicefrac{1}{2}$. For semi-empirical Hamiltonians,
TINC found that optimization of the Thouless functional $\omega[\vec{v}]=\vec{v}\cdot\mathbb{L}\cdot\vec{v}/\left|\vec{v}\cdotp\vec{v}\right|,$
corresponding to the solution of Eq.~(\ref{eq:RPA}), was significantly
slower for the RPA relative to the TDA, and also compared to subspace
solvers. For first principles models and non-trivial basis sets, this
naive RQI can become pathologically slow as shown in Fig.~\ref{fig:RPA_VS_TDA}.
On the other hand, the Tsiper formulation exposes the underlying pseudo-Hermitian
structure of the TD-SCF equations. Here, this structure is exploited
with QUasi-Independent Rayleigh Quotient Iteration (QUIRQI), involving
dual channel optimization of the Tsiper quotients coupled only weakly
through line search.

Our development begins with a brief review of the representation independent
formulation developed by TINC, which avoids the $\mathcal{O}(N^{3})$
cost of rotating into an explicit \emph{p-h}, \emph{h-p} symmetry.
Instead, this symmetry is maintained implicitly via annihilation,
$\boldsymbol{x}\leftarrow f_{a}(\boldsymbol{x})\boldsymbol{=P}\cdotp\boldsymbol{x}\cdotp\boldsymbol{Q}+\boldsymbol{Q}\cdotp\boldsymbol{x}\cdotp\boldsymbol{P}$,
with $\boldsymbol{P}$ the first order reduced density matrix and
$\boldsymbol{Q=I-P}$ its compliment. Likewise, the indefinite metric
associated with the $J$-symmetry of Eq.~(\ref{eq:RPA}) is carried
through the generalized norm $\left\langle \boldsymbol{x},\boldsymbol{y}\right\rangle ={\rm tr}$$\left\{ \boldsymbol{x}^{{\scriptscriptstyle T}}\cdotp\left[\boldsymbol{y},\boldsymbol{P}\right]\right\} $.
Introducing the operator equivalents, $\boldsymbol{L}[\boldsymbol{p}]\Leftrightarrow\mathbb{K}.\vec{p}$
and $\boldsymbol{L}[\boldsymbol{q}]\Leftrightarrow\mathbb{T}.\vec{q}$
, the Tsiper functional becomes $\omega\left[\boldsymbol{p},\boldsymbol{q}\right]=\frac{\left\langle \boldsymbol{p},\boldsymbol{L\left[p\right]}\right\rangle }{2\left|\left\langle \boldsymbol{p},\boldsymbol{q}\right\rangle \right|}+\frac{\left\langle \boldsymbol{q},\boldsymbol{L}[\boldsymbol{q}]\right\rangle }{2\left|\left\langle \boldsymbol{p},\boldsymbol{q}\right\rangle \right|}.$
Transformations between the transition density and the dual space
involves simple manipulations and minimal cost, allowing Fock builds
with the transition density and optimization in the dual space. The
splitting operation is given by $\boldsymbol{p}=f_{+}(\boldsymbol{v})=\boldsymbol{P}\cdot\boldsymbol{v}\cdot\boldsymbol{Q}+\left[\boldsymbol{Q}\cdot\boldsymbol{v}\cdot\boldsymbol{P}\right]^{{\scriptscriptstyle T}}$
and $\boldsymbol{q}=f_{-}(\boldsymbol{v})=\boldsymbol{P}\cdot\boldsymbol{v}\cdot\boldsymbol{Q}-\left[\boldsymbol{Q}\cdot\boldsymbol{v}\cdot\boldsymbol{P}\right]^{{\scriptscriptstyle T}}$.
Likewise, $\boldsymbol{L}[\boldsymbol{p}]=f_{-}\left(\boldsymbol{L}[\boldsymbol{v}]\right)$
and $\boldsymbol{L}[\boldsymbol{q}]=f_{+}\left(\boldsymbol{L}[\boldsymbol{v}]\right)$.
The back transformation (merge) from dual to density is $\boldsymbol{v}=F(\boldsymbol{p},\boldsymbol{q})=\left(\boldsymbol{p}+\boldsymbol{q}+\left[\boldsymbol{p-q}\right]^{{\scriptscriptstyle T}}\right)/2$.
This framework provides the freedom to work in any orthogonal representation,
and to switch between transition density and oscillator duals with
minimal cost.

QUIRQI is given in Algorithm \ref{quirqi}. It begins with a guess
for the transition density, which is then split into its dual (lines
2-3). The choice of initial guess is discussed later. Lines 4-24 consist
of the non-linear conjugate gradient optimization of the nearly independent
channels: In each step, the flow of information proceeds from optimization
of the duals to builds involving the density and back to the duals
in a merge-annihilate-truncate-build-split-truncate (MATBST) sequence.
For the variables $\boldsymbol{v}$, $\boldsymbol{p}$ and $\boldsymbol{q}$
this sequence is comprised by lines 22-23 and 5-7, and lines 15-19
for the corresponding conjugate gradients $\boldsymbol{h}_{v}$, $\boldsymbol{h}_{p}$
and $\boldsymbol{h}_{q}$. Truncation is carried out with the filter
operation as described in Ref.~\cite{ANiklasson03}, with cost and
error determined by the matrix threshold $\tau_{\mathrm{mtx}}$. 

The Tsiper functional is the sum of dual quotients $\omega_{p}$ and
$\omega_{q}$, determined at line 8, followed by the gradients $\boldsymbol{g}_{p}$
and $\boldsymbol{g}_{q}$ computed at line 9. After the first cycle,
the corresponding relative error $e_{\mathrm{rel}}$ (10) and maximum
matrix element of the gradient $g_{\mathrm{max}}$ (11) are computed
and used as an exit criterion at line 4, along with non-variational
behavior $\omega>\omega^{\mathrm{old}}.$

Next, the Polak-Ribiere variant of non-linear conjugate gradients
yields the search direction in each channel, $\boldsymbol{h}_{p}$
and $\boldsymbol{h}_{q}$ (12-14). The action of $\boldsymbol{L}[\cdot]$
on to $\boldsymbol{h}_{p}$ and $\boldsymbol{h}_{q}$ is then computed,
again with a MATBST sequence (15-19), followed by a self-consistent
dual channel line search at line 20, as described below. With steps
$\lambda_{p}$ and $\lambda_{q}$ in hand, minimizing updates are
taken along each conjugate direction (22), and the cycle repeats with
the MATBST sequence spanning lines 21-23 and 5-7.

\begin{algorithm}[h]
\caption{QUIRQI}

\label{quirqi} \begin{algorithmic}[1]

\Procedure{QUIRQI}{$\omega,\boldsymbol{v}$}

\State guess $\boldsymbol{v}$ 

\State$\boldsymbol{p}=f_{+}\left(\boldsymbol{v}\right)$, $\boldsymbol{q}=f_{-}\left(\boldsymbol{v}\right)$ 

\While{$e_{\mathrm{rel}}>\epsilon$ and $g_{\mathrm{max}}>\gamma$
{and $\omega<\omega^{\mathrm{old}}$}}

\State $\boldsymbol{L}[\boldsymbol{v}]=[\bm{F},\bm{\, v}]+[\bm{G}[\boldsymbol{v}],\,\bm{P}]$ 

\State$\boldsymbol{L}[\boldsymbol{p}]=f_{-}\left(\boldsymbol{L}[\boldsymbol{v}]\right)$,
$\boldsymbol{L}[\boldsymbol{q}]=f_{+}\left(\boldsymbol{L}[\boldsymbol{v}]\right)$ 

\State$\mathtt{\mathtt{\mathrm{\mathtt{filter}}}}\left(\boldsymbol{\boldsymbol{L}[\boldsymbol{p}]},\boldsymbol{\, L}[\boldsymbol{q}],\,\tau_{\mathrm{mtx}}\right)$

\State $\omega_{p}=\frac{\left\langle \mathbf{p},\boldsymbol{L}[\boldsymbol{p}]\right\rangle }{2\left|\left\langle \boldsymbol{p},\boldsymbol{q}\right\rangle \right|}$,
$\omega_{q}=\frac{\left\langle \mathbf{q},\boldsymbol{L}[\boldsymbol{q}]\right\rangle }{2\left|\left\langle \boldsymbol{p},\boldsymbol{q}\right\rangle \right|}$,
$\omega=\omega_{p}+\omega_{q}$

\State $\boldsymbol{g}_{p}=\boldsymbol{q\,}\omega_{q}-\boldsymbol{L}[\boldsymbol{p}]$,
$\boldsymbol{g}_{q}=\boldsymbol{p\,}\omega_{p}-\boldsymbol{L}[\boldsymbol{q}]$

\State $e_{\mathrm{rel}}=\left(\omega^{\mathrm{old}}-\omega\right)/\omega$

\State $g_{\mathrm{max}}=\underset{i,j}{\operatorname{max}}\left\{ \left[\boldsymbol{g}_{p}\right]_{ij},\left[\boldsymbol{g}_{p}\right]_{ij}\right\} $

\State $\beta_{p}=\frac{\left\langle \boldsymbol{g}_{p},\,\boldsymbol{g}_{p}-\boldsymbol{g}_{p}^{{\rm old}}\right\rangle }{\left\langle \boldsymbol{g}_{p}^{{\rm old}},\boldsymbol{g}_{p}^{{\rm old}}\right\rangle }$,
$\beta_{q}=\frac{\left\langle \boldsymbol{g}_{q},\,\boldsymbol{g}_{q}-\boldsymbol{g}_{q}^{{\rm old}}\right\rangle }{\left\langle \boldsymbol{g}_{q}^{{\rm old}},\boldsymbol{g}_{q}^{{\rm old}}\right\rangle }$

\State$\omega^{old}\leftarrow\omega$, $\boldsymbol{g}_{p}^{old}\leftarrow\boldsymbol{g}_{p}$
, $\boldsymbol{g}_{q}^{old}\leftarrow\boldsymbol{g}_{q}$ 

\State$\boldsymbol{h}_{p}\leftarrow\boldsymbol{g}_{p}+\beta_{p}\boldsymbol{h}_{p}$,
$\boldsymbol{h}_{q}\leftarrow\boldsymbol{g}_{q}+\beta_{q}\boldsymbol{h}_{q}^{{\rm }}$

\State$\boldsymbol{h}_{v}=F(\boldsymbol{h}_{p},\boldsymbol{h}_{q})$,
$\boldsymbol{h}_{v}\leftarrow f_{a}\left(\boldsymbol{h}_{v}\right)$

\State$\mathrm{\mathtt{filter}}\left(\boldsymbol{h}_{p},\boldsymbol{h}_{q},\boldsymbol{h}_{v},\,\tau_{\mathrm{mtx}}\right)$

\State$\boldsymbol{L}\left[\boldsymbol{h}_{v}\right]=[\bm{F},\,\boldsymbol{h}_{v}]+[\bm{G}[\boldsymbol{h}_{v}],\,\bm{P}]$

\State$\boldsymbol{L}[\boldsymbol{h}_{p}]=f_{-}\left(\boldsymbol{L}[\boldsymbol{h}_{v}]\right)$,
$\boldsymbol{L}[\boldsymbol{h}_{q}]=f_{+}\left(\boldsymbol{L}[\boldsymbol{h}_{v}]\right)$

\State$\mathrm{\mathtt{filter}}\left(\boldsymbol{L}[\boldsymbol{h}_{p}],\,\boldsymbol{L}[\boldsymbol{h}_{q}],\,\tau_{\mathrm{mtx}}\right)$

\State$\left\{ \lambda_{p},\lambda_{q}\right\} =\underset{\left\{ \lambda_{p},\lambda_{q}\right\} }{\operatorname{argmin}}\;\omega\left[\boldsymbol{p}+\lambda_{p}\boldsymbol{h}_{p},\boldsymbol{q}+\lambda_{q}\boldsymbol{h}_{q}\right]$

\State$\boldsymbol{p}\leftarrow\boldsymbol{p}+\lambda_{p}\boldsymbol{h}_{p}$,
$\boldsymbol{q}\leftarrow\boldsymbol{q}+\lambda_{q}\boldsymbol{h}_{q}$

\State$\boldsymbol{v}\leftarrow F\left(\boldsymbol{p},\boldsymbol{q}\right)$,
$\boldsymbol{v}\leftarrow f_{a}\left(\boldsymbol{v}\right)$

\State$\mathrm{\mathtt{filter}}\left(\boldsymbol{p},\boldsymbol{q},\boldsymbol{v},\,\tau_{\mathrm{mtx}}\right)$

\EndWhile\label{euclidendwhile-1} 

\EndProcedure

\end{algorithmic} \label{Flo:QUIRQI}
\end{algorithm}

Optimization of the Tsipper functional $\omega\left[\lambda_{p},\lambda_{q}\right]\equiv\omega\left[\boldsymbol{p}+\lambda_{p}\mathbf{h}_{p},\boldsymbol{q}+\lambda_{q}\boldsymbol{h}_{q}\right]$
involves a two dimensional line-search (line 20) corresponding to
minimization of \begin{equation}
\omega\left[\lambda_{p},\lambda_{q}\right]=\frac{A_{p}+\lambda_{p}B_{p}+\lambda_{p}^{2}C_{p}+A_{q}+\lambda_{q}B_{q}+\lambda_{q}^{2}C_{q}}{R_{pq}+\lambda_{p}S_{pq}+\lambda_{q}T_{pq}+\lambda_{p}\lambda_{q}U_{pq}}\:,\label{eq:2DLineSearch}\end{equation}
with coupling entering through terms in the denominator such as $U_{pq}=\left\langle \boldsymbol{h}_{p},\boldsymbol{h}_{q}\right\rangle $.
A minimum in Eq.~(\ref{eq:2DLineSearch}) can be found quickly to
high precision by alternately substituting one-dimensional solutions
one into the other until self-consistency is reached. This semi-analytic
approach starts with a rough guess at the pair $\left\{ \lambda_{p},\lambda_{q}\right\} $
(eg. found by a coarse scan) followed by iterative substitution, where
for example the $p$-channel update is 

\begin{eqnarray}
\lambda_{p}\leftarrow\left\{ \left[\left(2C_{p}R_{pq}+2C_{p}\lambda_{q}S_{pq}\right)^{2}-4\left(C_{p}T_{pq}+C_{p}\lambda_{q}U_{pq}\right)\right.\right.\nonumber \\
\left[B_{p}R_{pq}+B_{p}\lambda_{q}S_{pq}-\left(A_{q}+A_{p}+B_{q}\lambda_{q}+C_{q}\lambda_{q}^{2}\right)T_{pq}\right.\nonumber \\
\left.\vphantom{\left[\left(2C_{q}R_{pq}+2C_{q}\lambda_{p}S_{pq}\right)^{2}-4\left(C_{q}T_{pq}+C_{q}\lambda_{p}U_{pq}\right)\right.}\left.-\left(A_{q}\lambda_{q}+A_{q}\lambda_{q}+B_{q}\lambda_{q}^{2}+C_{q}\lambda_{q}^{3}\right)U_{pq}\right]\right]^{1/2}\qquad\label{eq:StepLengthP}\\
\left.\vphantom{\left\{ \left[\left(2C_{q}R_{pq}+2C_{q}\lambda_{p}S_{pq}\right)^{2}-4\left(C_{q}T_{pq}+C_{q}\lambda_{p}U_{pq}\right)\right.\right.}-2C_{p}R_{pq}-2C_{p}\lambda_{q}S_{pq}\right\} \big/\left[2C_{p}\left(T_{pq}+\lambda_{q}U_{pq}\right)\right]\,,\qquad\nonumber \end{eqnarray}
with an analogous update for the $q$-channel obtained by swapping
subscripts. As the solution decouples ($S_{pq},$ $T_{pq}$ and $U_{pq}$
become small) the steps are found independently. 

QUIRQI has been implemented in FreeON \cite{FreeON}, which employs
the linear scaling Coulomb and Hartree-Fock exchange kernels QCTC
and ONX with cost and accuracy controlled by the two-electron screening
threshold $\tau_{\mathrm{2e}}$ \cite{ESchwegler97}. \emph{N}-scaling
solution of the QUIRQI matrix equations is achieved with the sparse
approximate matrix-matrix multiply (SpAMM), with cost and accuracy
determined by the drop tolerance $\tau_{\mathrm{mtx}}$ \cite{MChallacombe99,MChallacombe00B,ANiklasson03}.
All calculations were carried out with version 4.3 of the gcc/gfortran
compiler under version 8.04 of the Ubuntu Linux distribution and run
on a fully loaded 2GHz AMD Quad Opteron 8350.

For systems studied to date, QUIRQI is found to converge monotonically
with rates comparable to the TDA as shown in Fig.~\ref{fig:RPA_VS_TDA}.
Based on the comparative performance presented by TINC, the TDA rate
of convergence appears to be a lower bound for RPA solvers. In addition
to the convergence rate, performance is strongly determined by the
initial guess. The following results have been obtained using the
polarization response density along the polymer axis \cite{STretiak09},
which can be computed in $\mathcal{O}(N)$ by Perturbed Projection
\cite{VWeber04}. Also, a relative precision of 4 digits in the excitation
energy is targeted with the convergence parameters $\epsilon=10^{-4}$
and $\gamma=10^{-3}$, with exit from the optimization loop on violation
of monotonic convergence ($\omega>\omega^{\mathrm{old}}$ due to precision
limitations associated with linear scaling approximations).

\begin{figure}[h]
\includegraphics[width=3.3in]{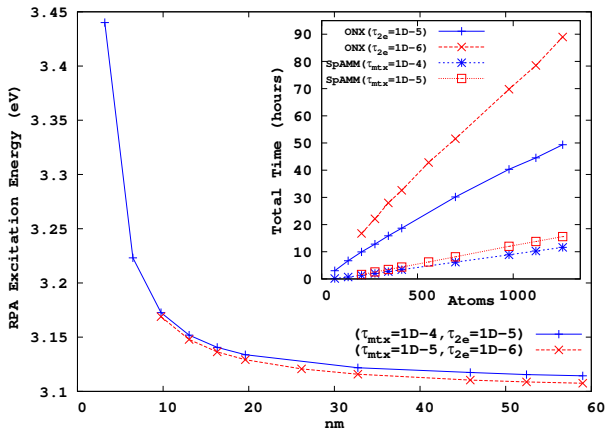}

\caption{Approach to the bulk limit of the PPV first excited state at the 6-31G{*}{*}/RPA
level of theory, with inset showing linear scaling cost for HF exchange
(ONX) and sparse linear algebra (SpAMM). The cost of Coulomb sums
with much tighter thresholds are comparable to those for the SpAMM.
\label{fig:PPVScaling}}

\end{figure}

\begin{figure}[h]
\includegraphics[width=3.3in]{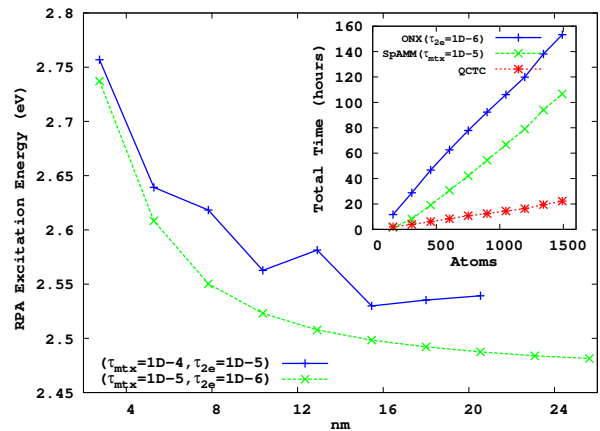}

\caption{Approach to the bulk limit of the first excited state of the (4,3)
carbon nanotube segment at the 3-21G/RPA level of theory, with inset
showing linear scaling cost for HF exchange (ONX), sparse linear algebra
(SpAMM) and Coulomb sums (QCTC). \label{fig:NTScaling}}

\end{figure}

In Fig.~\ref{fig:PPVScaling}, linear scaling and convergence to
the bulk limit are demonstrated for a series of polyphenylene vinylene
(PPV) oligomers at the RHF/6-31G{*}{*} level of theory for the threshold
combinations $\left\{ \tau{}_{\mathrm{mtx}},\tau_{\mathrm{2e}}\right\} =\left\{ 10^{-4},10^{-5}\right\} $
and $\left\{ 10^{-5},10^{-6}\right\} $. Significantly more conservative
thresholds have been used for the Coulomb sums, which incur only minor
cost. Convergence is reached in $24-25$ iterations, with the cost
of Coulomb summation via QCTC comparable to the cost of SpAMM($\tau_{\mathrm{mtx}}=10^{-4}$).
In Fig.~\ref{fig:NTScaling}, linear scaling and convergence to the
bulk limit are demonstrated for a series of (4,3) carbon nanotube
segments at the RHF/3-21G level of theory for the same threshold combinations,
again with convergence achieved in about 24-25 cycles. In both cases,
tightening the pair $\left\{ \tau_{\mathrm{mtx}},\tau_{\mathrm{2e}}\right\} $
leads to a systematically improved result. While the $\left\{ 10^{-4},10^{-5}\right\} $
thresholds that work well for PPV lead to a non-monotone behavior
with respect to extent for the nanotube series, dropping one more
decade to $\left\{ 10^{-5},10^{-6}\right\} $ leads to a sharply improved
behavior. Dropping thresholds further to $\left\{ 10^{-6},10^{-7}\right\} $
yields identical results to within the convergence criteria ($\sim$~four
digits) across the series, also scaling with $N$ but at roughly twice
the cost. 

These results demonstrate that QUIRQI can achieve both systematic
error control and linear scaling in solution of the RPA eigenproblem
for systems with extended conjugation. Relative to PPV, the greater
numerical sensitivity encountered with the nanotube series is consistent
with the ground state problem, where a smaller band gap and greater
atomic connectivity typically demand tighter thresholds. 

QUIRQI exploits decoupling of the Tsipper functional into nearly independent,
pseudo-Hermitian quotients leading to aggressive convergence rates
equivalent to the fully Hermitian TDA, while remaining variational
with respect to matrix truncation ($\tau_{\mathrm{mtx}}$). However,
QUIRQI is not variational with respect to the screening parameter
$\tau_{\mathrm{2e}}$. It can be systematically improved by tightening
$\tau_{\mathrm{2e}}$ though, due to rigorous error bounds based on
the Schwartz inequality \cite{ESchwegler97}. These properties present
opportunities for more precise error control as suggested by Rubensson,
Rudberg and Salek \cite{ERubensson08}. Further, these properties
are expected to hold even for the most general SCF models, with the
only difference being an increasingly localized transition density
matrix and larger cost prefactor with an increasing DFT component.
Finally, a variational approach allows considerable flexibility in
the path to solution, as errors due to approximation can be overcome
by optimization, offering opportunities for single precision GPU acceleration,
variable thresholding, incremental Fock builds as well as extrapolation
techniques. 

This work was supported by the U.S. Department of Energy and Los Alamos
LDRD funds. Los Alamos National Laboratory was operated by the Los
Alamos National Security, LLC, for the National Nuclear Security Administration
of the U.S. Department of Energy under Contract No. DE-AC52-06NA25396.
Special acknowledgments go the International Ten Bar Caf{\'e} for scientific
vlibations, and to Sergei Tretiak and Nicolas Bock for valuable input.


\end{document}